\documentclass{ws-ijmpa}

\begin{document}

\title{Summary of Experimental Meson Physics}

\author{Kamal K. Seth}

\address{Northwestern University, Evanston, Illinois, IL 60208, USA\\kseth@northwestern.edu}

\maketitle

\begin{abstract}
A summary of the present experimental status of meson physics is presented.  The presentation includes the new results presented at the MESON06 workshop, as well as other recent experimental developments in the field.
\end{abstract}

\ccode{PACS numbers: 14.40.-n,13.25.-k,25.80.-e}

\section{Introduction}

Meson physics is a very broad field, and this workshop covers a very large part of it, with about 25 plenary and 50 parallel talks.  I can not possible attempt to summarize all that.  So, I will pick and choose from what was presented, a process which is admittedly very subjective, and add some new and interesting results which were not talked about.  I will also update some results which have appeared in print since my talk.

Before I go into physics, let me acknowledge the laboratories and experiments which have contributed to this field, and to this workshop, and briefly mention what is, to the best of my knowledge, their current status.  In the following, `ended' means that data taking has ended, but physics analysis and publications continue. 
\begin{quote}
BEPC (BES II ended, BES III operational in 2007?); BNL (E852 ended, PHOENIX and STAR at RHIC have minor spectroscopy programs); CELSIUS (WASA ended); CERN (Crystal Barrel and OBELIX ended; COMPASS has potential); COSY (ANKE, GEM, TOF operational); DESY (H1, ZEUS, HERMES operational), Fermilab (SELEX and FOCUS ended, Tevatron operational until 2009), Frascati (at DA$\Phi$NE; KLOE and DEAR ended, FINUDA and SIDDHARTA operational); GSI (HADES operational, PANDA in 2014?); JLab (CLAS operational, 12 GeV upgrade?); JPARC (operational in 2008?), KEK (PS and Belle (high potential) operational; Bonn (ELSA operational); Mainz (MAMI II operational); SLAC (BaBar (high potential) operational until 2010).
\end{quote}
The sum total of the above summary is that while of the many very productive older programs, some have ended, many continue, and new programs (BEPC II, GSI (FAIR), and JPARC) are in the offing.  We can therefore have high expectations for continued exciting discoveries in meson physics in the future.

I will begin my summary in the order of the mass hierarchy.  Light quark ($n (=u,d)~\mathrm{and}~s$) mesons, heavy quark ($c,b$) mesons, and then light/heavy ($c/n,b/n$) mesons.  For good measure, I will throw in a gluon or two, and say a few words about glueballs.  Finally, I will talk about the recently discovered exotics, about which everybody is excited, but whose nature nobody really knows.

\section{Pions and Kaons}

Pions and kaons are ancient particles, and one would think that there could not possible by anything new to learn about them.  Actually, there is!  One of the most interesting and elementary questions one can ask about a particle is ``what does it look like?''  To put it in terms of the simplest physical observable, the question becomes, ``what do the particle's electromagnetic form factors look like?''  To relate the meson's form factors to its constituent quarks one needs to invoke perturbative QCD (at least at present), and that means that one needs experimental measurements of form factors at large enough momentum transfers at which pQCD might be expected to be valid.  Let us see where we stand both experimentally and theoretically on this issue.

\begin{figure}[!tb]
\begin{center}
\includegraphics[width=2.4in]{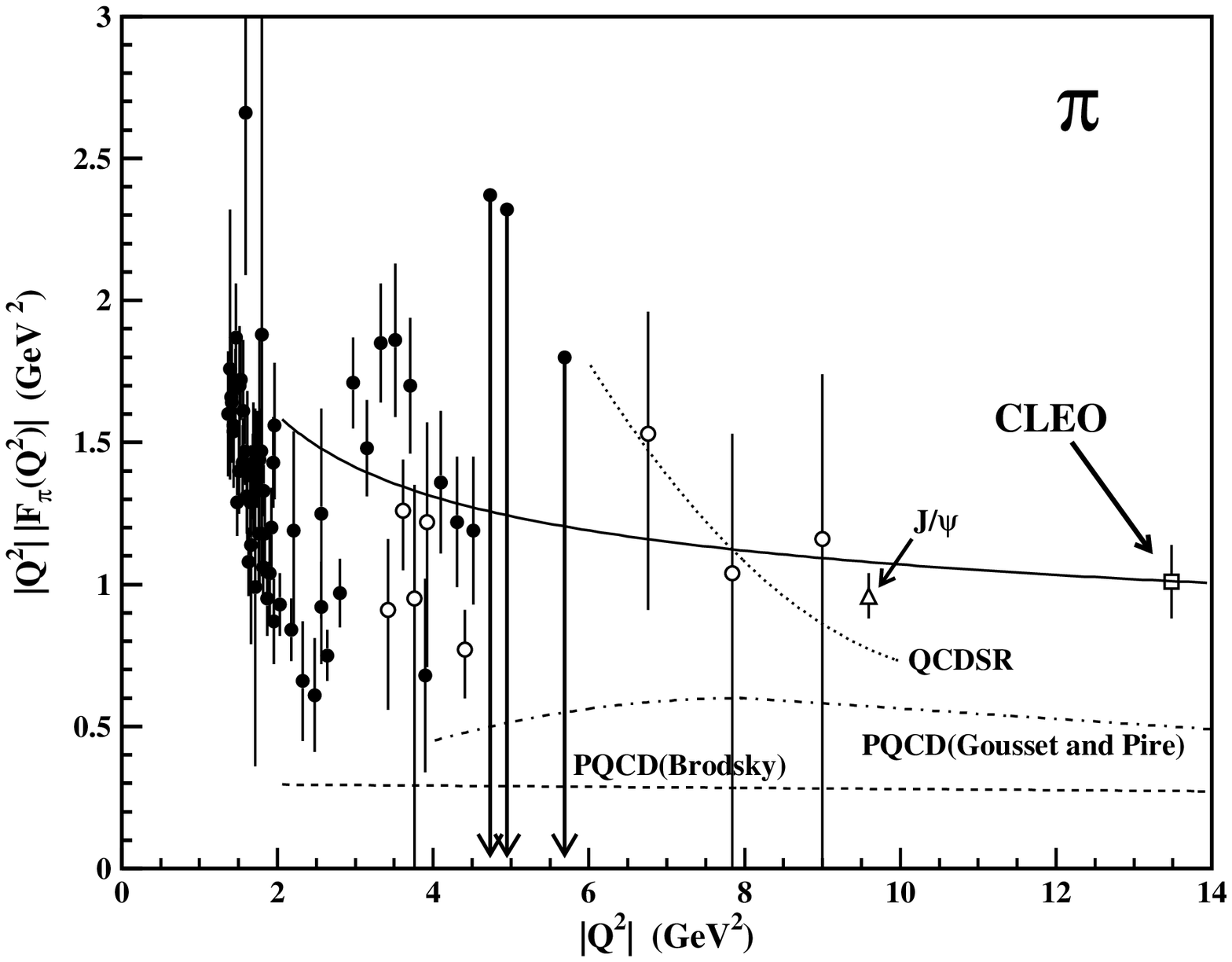}
\includegraphics[width=2.4in]{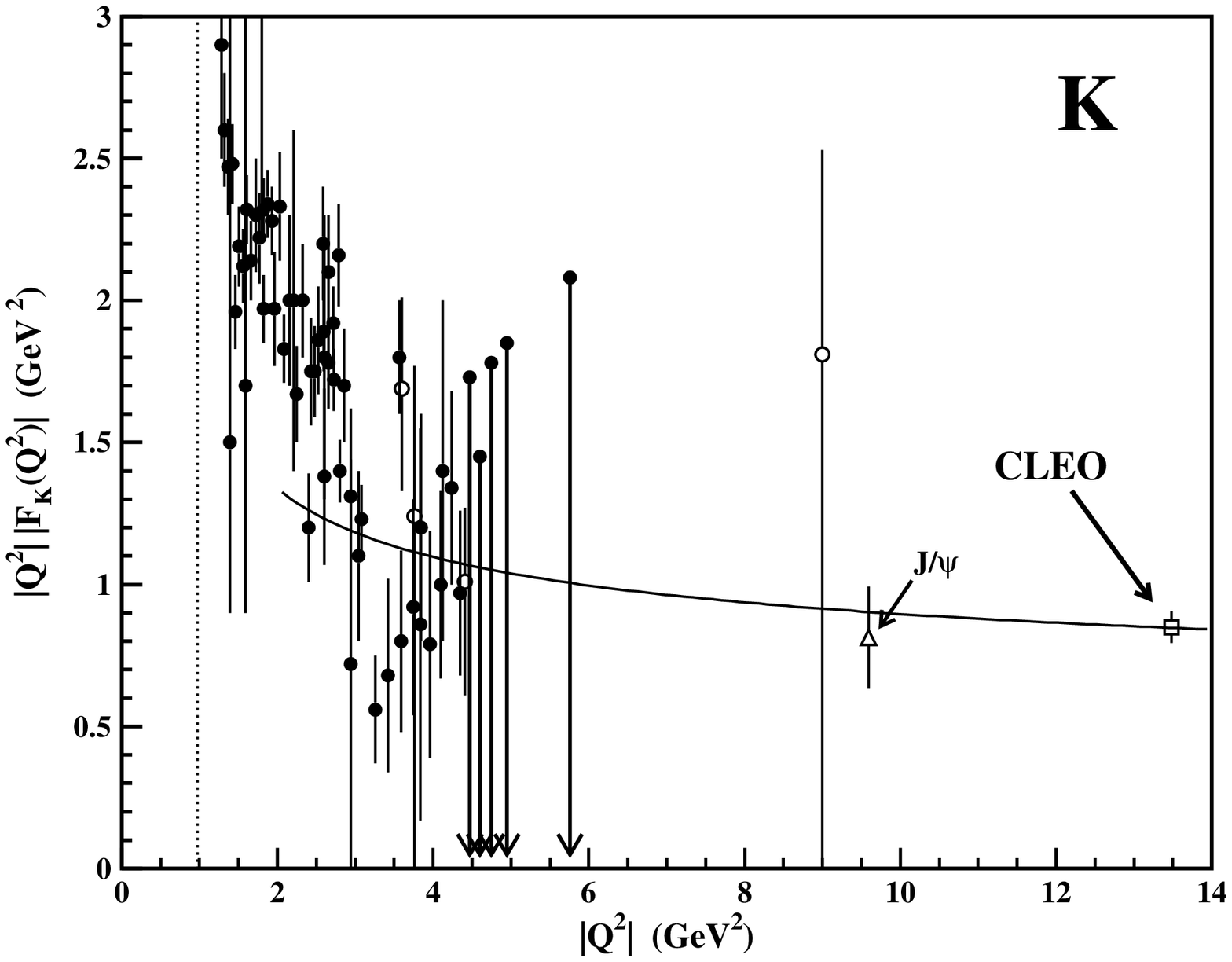}
\end{center}
\caption{Timelike electromagnetic form factors of pion and kaon.  The new CLEO results for $|Q^2|=13.48~\mathrm{GeV}^2$ are shown with square symbols.}
\end{figure}

\subsection{Electromagnetic Form Factors of Pions and Kaons}

Let me recall the great debate about ``when is $Q^2$ large enough to validate pQCD?''  which took place in the mid-1980s between Brodsky and colleagues\cite{lepagebrodsky}, who believed $Q^2\approx10~\mathrm{GeV}^2$ was enough, and Isgur and Llwellyn Smith\cite{isgursmith}, who claimed that $Q^2>100~\mathrm{GeV}^2$ would be required.  The experimental data for pion form factors used in the debate consisted of barely three data points for $Q^2>2~\mathrm{GeV}^2$  with $\pm50\%$ errors.  We would all agree that perhaps this was ``much ado about nothing''.  So, what is new?

What is new is the first precision measurements at large timelike momentum transfer ($Q^2=-s\approx13.48~\mathrm{GeV}^2$) of pion and kaon form factors at CLEO\cite{cleoff} as shown in Fig. 1\footnote{The $Q^2F_\pi$ point marked ``$J/\psi$'' is not a direct measurement, but dervied from the measured branching ratio for $J/\psi\to\pi^+\pi^-$ by Milana et al.\cite{milana}.  Using the same arguments, I have derived $Q^2F_K$\cite{seth}.  The agreement of these $Q^2F_{\pi,K}$ at $Q^2=9.6~\mathrm{GeV}^2$ with the measured values at $Q^2=13.5~\mathrm{GeV}^2$ lends justification to the procedure.}.  The solid curves in the figure show the arbitrarily normalized variation of $\alpha_{\mathrm{Strong}}$ with $Q^2$.  Also shown for the pion are pQCD predictions and QCD sum rule predictions, neither of which come even close to the measured values, which says that here is now something real for theorists to chew on.

\section{The S--Wave Mesons}

 Most of the activity in this category has been confined to the pseudoscalar $\eta$ and $\eta'$ mesons.  The obvious reason is that accelerators in the $1-3$ GeV region can produce them copiously.  At this conference, a lot of activity was reported on production of $\eta$ and $\eta'$.  Detailed reports were presented on photo-- and electro--production measurements being made at Mainz, JLab, and ELSA.  We also heard about pion-induced production at BNL, production in $e^+e^-$ collisions at Frascati, in $pp$ collisions at Julich, and in heavy ion collisions at GSI.

We have an interesting development in the measurements of the mass of $\eta$.  A recent COSY measurement\cite{cosyetamass} reported $M(\eta)=547.311\pm0.028(\mathrm{stat})\pm0.032(\mathrm{syst})~\mathrm{MeV}$, which differs by $\sim8\sigma$ from an earlier NA38 measurment\cite{na38etamass}. $M(\eta)=547.843.0.030(\mathrm{stat})\pm0.041(\mathrm{syst})$ MeV.  Now we have the report from KLOE\cite{kloeetamass} of $M(\eta)=547.822\pm0.005(\mathrm{stat})\pm0.069(\mathrm{syst})$, which agrees with NA38 within $0.25\sigma$.  The resolution of the $\sim500$ keV difference in $M(\eta)$ between COSY and NA38/KLOE would normally be only of academic interest, but in this era of meson--meson molecules, it could be important if somebody comes up with a weakly bound molecule of $(\eta+X)$!

The mass of $\eta'$ is known with a $\pm140$ keV uncertainty\cite{pdg}, but so far we have no reports from COSY or KLOE on improved measurements.

KLOE\cite{kloeetamass} has also reported the long--sought value of the ratio $R\equiv\mathcal{B}(\phi\to\gamma\eta')/\mathcal{B}(\phi\to\gamma\eta)=(4.74\pm0.09\pm0.20)\times10^{-3}$, and used it to address the question of the possible glue content of $\eta'$.  They find it to be $1-(0.92\pm0.06)=(8\pm6)\%$.  While the numerical value is not significant at this level, it does indicate that the glue content of $\eta'$ is quite small.

The spectroscopy of the low--mass vectors, the $\rho,~\omega,~\phi$ has been dominated by the precision measurements of the Novosibirsk SND and CMD2 groups.  Measurements of comparable precision are now being reported by KLOE.  Hopefully, long--standing problems, like the $>5$ MeV uncertainty in the $\rho$ mass can be resolved by these new data, although it is my belief that the problem actually lies with the different formalisms for resonance analysis being used.

\section{The P--Wave Mesons}

The P--wave singlets, $h_1$ and $b_1$, have remained inactive for the last fifteen years.  The focus of the activity has been on the P--wave triplets, and among them the greatest attention has been drawn by the isosinglet scalars, primarily because $I=0,~J^{PC}=0^{++}$ are also the quantum numbers of the lowest mass glueball.  It behooves us, however, to take a look at all three, $^3P_0(0^{++})$, $^3P_1(1^{++})$, $^3P_2(2^{++})$.

The problem common to all the states with masses larger than $\sim1$ GeV is excessive crowding.  In the 1 GeV interval of mass, $1.25-2.25$ GeV, according to Godfrey and Isgur\cite{godfreyisgur}, 75 ($u,d,s$) mesons are expected, the average spacing being $\sim15$ MeV.  The ones which have been observed have widths ranging from $100-500$ MeV, which means that states overlap, mix, and interfere horribly, and even the most sophisticated partial width analyses of experimental data have a tough time identifying and characterizing (i.e. $J^{PC}$, branching ratios) the states.  The problem is best illustrated by the isospin singlet scalars, the $0^{++}~f_0$'s.

\subsection{The Isosinglet $^3P_0(0^{++})$ or $f_0$ States}

With $u,d,s$ quarks, one expects two isoscalars, $f_0$ and $f_0'$, and an isovector $a_0$.  The trouble is that there are \textit{five} $f_0$'s known:  $f_0(600)$ or $\sigma$, $f_0(980)$, $f_0(1370)$, $f_0(1500)$, and $f_0(1710)$.  It has become very important to examine these very carefully because one expects only one more isosinglet $0^{++}$ state beyond $f_0$ and $f_0'$, and that third state is to be identified with the interesting, and ever elusive, glueball.  But we have five!  So, any new experimental information on these is most welcome.

\subsubsection{The Sigma, or $f_0(600)$: non--$q\bar{q}$ (?)}

The sigma has been there for a long time, albeit in the shadows.  Its most recent observation is by BES in $J/\psi\to\omega\sigma,~\sigma\to\pi^+\pi^-$, with the result $M(\sigma)=541\pm35~\mathrm{MeV}$, $\Gamma(\sigma)=502\pm84~\mathrm{MeV}$\cite{bessigma}.  Leutwyler and colleagues\cite{leutwyler} interpret the $\sigma$ as a pole in the $\pi\pi$ scattering amplitude and obtain $M(\sigma)=441^{+16}_{-8}~\mathrm{MeV}$, $\Gamma(\sigma)=544^{+18}_{-15}~\mathrm{MeV}$, and consider its physics to be governed by the Goldstone pions, i.e., the sigma is not a simple $q\bar{q}$ meson.  This is consistent with the fact that typical potential model calculations (e.g. Godfrey and Isgur\cite{godfreyisgur}) predict the lowest mass $f_0$ to have a mass of $\sim1.09$ GeV.  Lattice has so far nothing to say about the $\sigma$; it is still struggling with 250 MeV pions.  The fact is that the nature of the $\sigma$ remains as cloudy as ever, althought Pennington\cite{pennington} claims that its two photon width precludes its being a tetraquark ($\bar{q}\bar{q}qq$) or a glueball.

Parenthetically, let me mention that BES\cite{beskappa} has confirmed the observation of sigma's strange partner, the \textbf{kappa}, previously reported by Fermilab E791\cite{e791kappa} in the reaction $D^+\to(K^-\pi^+)\pi^+$ with $M(\kappa)=797\pm19\pm43~\mathrm{MeV}$ and $\Gamma(\kappa)=410\pm43\pm87~\mathrm{MeV}$.  In the reaction $J/\psi\to\overline{K^*}(890)+\kappa,~\kappa\to K^+\pi^-$, BES obtains $M(\kappa)=878\pm23^{+64}_{-38}~\mathrm{MeV}$ and $\Gamma(\kappa)=499\pm52^{+55}_{-87}~\mathrm{MeV}$.  It will be interesting to see if the $\kappa$ also emerges from the analysis of $K\pi$ scattering length, a la Leutwyler.

\subsubsection{The $f_0(980)$: non--$q\bar{q}$ (?)}

We had five scalars, and we wanted to weed out two in order to have left two $|q\bar{q}>$ and one glueball.  Well, we appear to have put $f_0(600)$ or $\sigma$ aside as a pole in the $\pi\pi$ scattering amplitude.  We need to remove one more.  The axe usually falls on $f_0(980)$.  Because $M(K\bar{K})=985$ MeV, and $f_0(980)$ strongly couples to $K\bar{K}$, $f_0(980)$, and its cousin $a_0(980)$, are often dismissed as $K\bar{K}$ molecules.  This conjecture, or, for example, the recent conjecture\cite{beveren} that like the $\sigma$, $f_0(980)$ is a dynamically generated resonance, make $f_0(980)$ non--$|q\bar{q}>$.  However, there is no hard proof, theoretical or experimental, for this assertion.  As a matter of fact, as described later, I believe that the systematics of $^3P_J$ states argues against this prejudice.

Despite the fact that $f_0(980)$ is one of the oldest known (since 1973) resonances, and has been observed in numerous experiments, very little quantitative is known about it.  An average of the latest measurements by E791\cite{e791f0}, Belle\cite{bellef0}, BES\cite{besf0}, and KLOE\cite{kloef0} gives $M(f_0(980))=979\pm2$ MeV, and the average of the E791\cite{e791f0} and Belle\cite{bellef0} results gives $\Gamma(f_0(980))=45\pm5~\mathrm{MeV}$.  Besides this, we know very little.  PDG06 summarizes the decays of $f_0(980)$ to $2\pi$, $2K$, and $2\gamma$ as \textit{seen}, \textit{seen}, \textit{seen}, which can be seen as a very unsatisfactory situation.  The only quantitative measure we have is that according to BES\cite{besf0}, the partial width for $K^+K^-$ is about $1/3$ that for $\pi^+\pi^-$, and when the relative phase space differences are taken into account, the coupling to kaons is about 3 to 4 times larger than that to pions\cite{besf0,kloef0}.  


\subsubsection{The $f_0(1370)$, $f_0(1500)$, and $f_0(1710)$: Glueballs, anyone?}

With $f_0(600)$ and $f_0(980)$ banished as non--$|q\bar{q}>$, the scalar field is reduced to the required three, two $|q\bar{q}>$ and one $|gg>$ glueball.  Most of the glueball games are played between these three.  Despite early over--optimistic searches for `pure' glueballs, it is now generally accepted that `pure' glueballs do not exist.  Quantum mechanics dictates that the scalar glueball must mix with the scalar mesons.  So, the game now is \textit{how much} $|gg>$ exists in each of the three candidates.  Is there a winner?  Perhaps not by a landslide, but even so!  The answer depends on who you ask, and how they order the \textit{unmixed} $|n\bar{n}>$, $|s\bar{s}>$ and $|gg>$ states.  Not much new has been forthcoming in this debate which used to be the talk of the town just a few years ago.  The $f_0(1710)$ is now firmly established as being $J^{PC}=0^{++}$, and it has been found to be healthily populated, along with $f_0(1500)$ in $\bar{p}p$ annihilation at 5.2 GeV/$c$\cite{uman}.

\begin{figure}[!tb]
\begin{center}
\includegraphics[width=4.in]{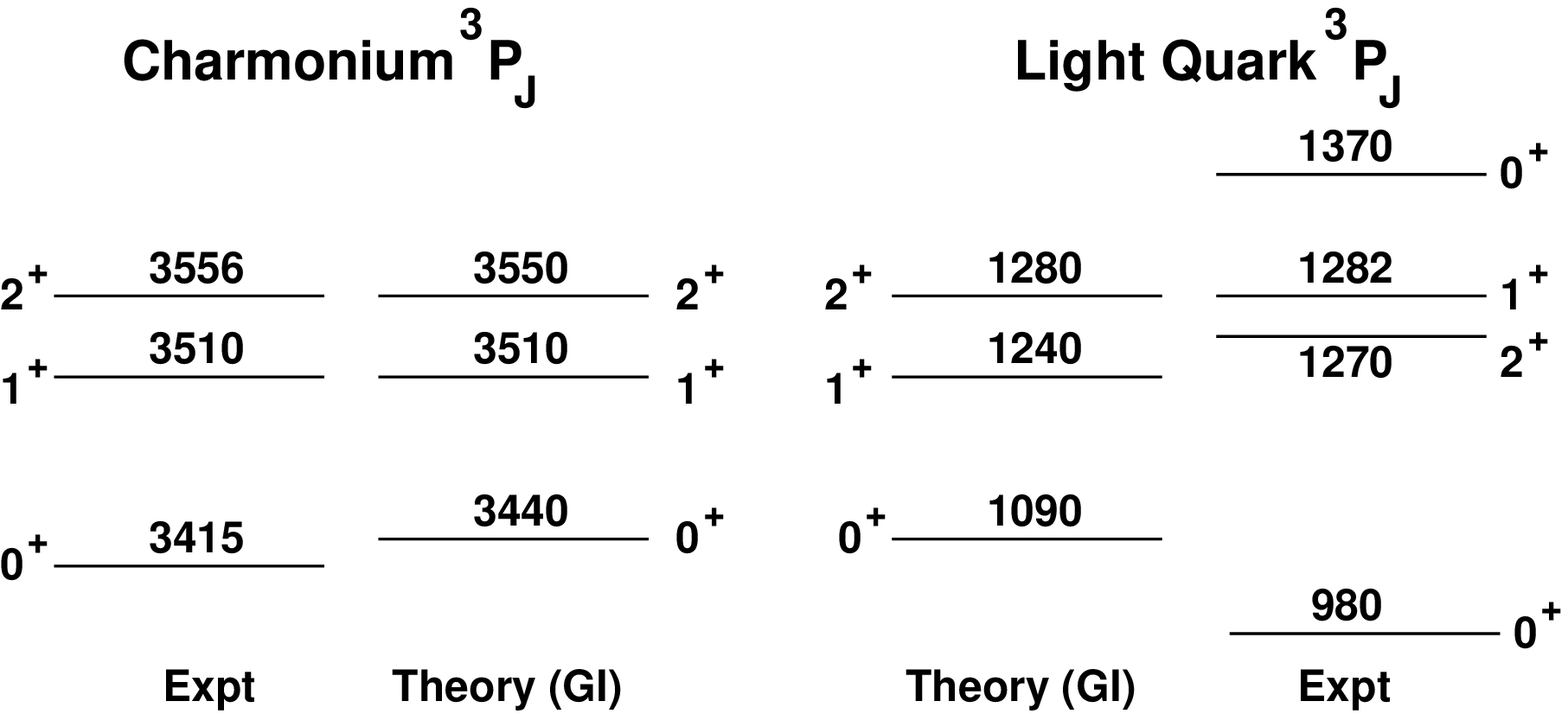}
\end{center}
\caption{Systematics of $^3P_J$ levels, (left) $|c\bar{c}>$ charmonium $^3P_J$ states in experiment and theory, (right) low--lying light quark $^3P_J$ states.}
\end{figure}

\subsection{The isoscalar $^3P_J$ and the $\vec{l}\cdot\vec{s}$ problem}

There is a vexing problem with the spin--orbit splitting of the lowest light quark $^3P_J$ states, and I believe that it relates to the non--$|q\bar{q}>$ assignment of $f_0(980)$.  Let me elaborate.

The QCD interaction does not depend on quark flavor.  The spin--orbit splitting has the same heirarchy for $|n\bar{n}>$, $|c\bar{c}>$ and $|b\bar{b}>$, and it is in the order: $0^{++}$ (lowest), $1^{++}$, $2^{++}$.  As shown in Fig.~2, it is the order predicted for, and experimentally observed for $|c\bar{c}>$ charmonium, and $|b\bar{b}>$ bottomonium.  It is what is predicted for  $|n\bar{n}>$, for example by Godfrey and Isgur\cite{godfreyisgur}.  In assigning $f_0(1370)$ as the $0^{+}$ member of the lowest $^3P_J$, the PDG would have us believe in the completely inverted order, $2^{++}$ (lowest), $1^{++}$, $0^{++}$.  This, of course, becomes necessary because the $f_0(980)$ has been dismissed as a non--$|q\bar{q}>$.  It is far easier for me to believe that the theoretically predicted $0^{++}$ at 1090 MeV has been shifted down to 980 MeV by non--$|q\bar{q}>$ admixtures, than to believe that it moves up by $\sim300$ MeV to 1370 MeV. In other words, I want to keep alive the possibility of $f_0(980)$ having a substantial component of $|q\bar{q}>$.  In doing so, I am entirely aware of the width arguments advanced by Godfrey and Isgur against $f_0(980)$ being $|q\bar{q}>$.

Having dwelt so long on the $0^{++}$ light--quark mesons, let me simply mention that the situation about $2^{++}$ tensors is no better.  Below 2.4 GeV in mass, six $f_2$ are predicted by Godfrey and Isgur\cite{godfreyisgur}, twelve are claimed by different experiments, and of these only six are considered as well established\cite{pdg}.

\begin{figure}[!tb]
\begin{center}
\includegraphics[width=3.in]{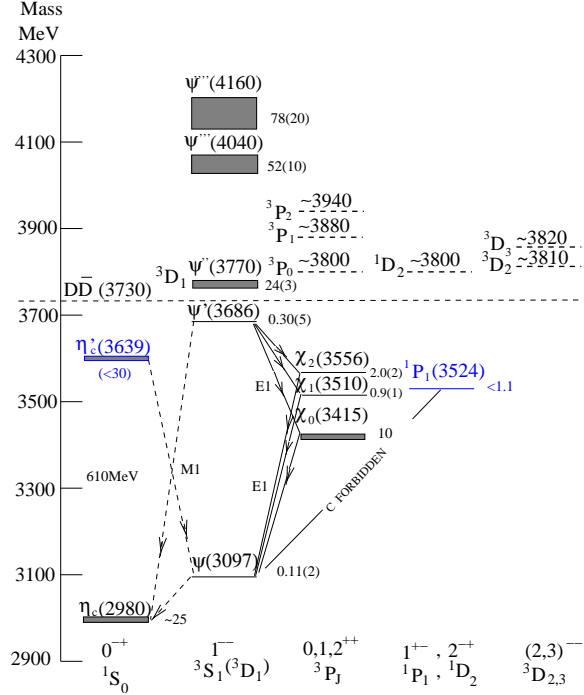}
\end{center}
\caption{Spectrum of $|c\bar{c}>$ charmonium states.  The dashed lines indicate states predicted by Godfrey and Isgur\protect\cite{godfreyisgur}.}
\end{figure}

\section{The Heavy--Quark Mesons}

As stated earlier, the light--quark mesons, have the experimental problem of numerous closely spaced overlapping states with very large widths, and the theoretical problems of being highly relativistic and having too large a value of $\alpha(\mathrm{strong})$ ($\gtrsim0.6$) to allow application of pQCD.  In contrast, the heavy quark mesons are far more tractable, both because they have well--resolved, narrow--width bound states (e.g., for $|c\bar{c}>$ typical level spacing and width are 100 MeV and $<10$ MeV, respectively), and tractable relativistic problems ($(v/c)^2\approx0.2~\mathrm{for}~|c\bar{c}>$, $0.1~\mathrm{for}~|b\bar{b}>$), and small enough $\alpha(\mathrm{strong})(\approx0.3~\mathrm{for}~|c\bar{c}>,$ $\approx0.2~\mathrm{for}~|b\bar{b}>$) to permit use of pQCD.  It follows that if you want to relate meson spectroscopy to the fundamental theory of QCD, you must turn to heavy--quark mesons, and this series of workshops must evolve more and more in that direction.

Let me begin by drawing your attention to the level diagram for $|c\bar{c}>$ charmonium states (in Fig.~3).  There are eight bound states before the system becomes unstable with respect to decay into open charm $D\bar{D}$ mesons ($D\equiv|c\bar{n}>$) at 3.73 GeV.  The $|b\bar{b}>$ bottomonium system is similar, but because of the three--times heavier mass of the $b$--quark ($m_b\approx4.5~\mathrm{GeV}$, $m_c\approx1.5~\mathrm{GeV}$), it has fourteen bound states before it becomes unstable with respect to decay into open--beauty $B\bar{B}$ mesons ($B\equiv|b\bar{n}>$) at 10.5 GeV.

Let me talk about the experimental situation. The bad news is that many of the facilities which produced most of the recent high precision hadron spectroscopy data have ended.  This includes: BNL (E852), CERN (CBarrel), FERMILAB (E760/E835). The good news is that new and promising facilities are just coming on, or are on the horizon.  These include:  DAFNE, GSI (FAIR), JPARC, BEPC II, and CEBAF Upgrade. The (unexpected) good news is that the SLAC and KEK $B$ factories, not built for spectroscopy, have accumulated such huge luminosities that they seem to be capable of making excellent contributions to hadron spectroscopy, almost wherever they choose to.

Progress is made as much by improved precision as by discoveries.  So, let me give you a few examples of precision measurements in heavy quark meson spectroscopy.  At Novosibirsk, the mass of $J/\psi$, $M(J/\psi)=3096.917\pm0.02~\mathrm{MeV}$, has been measured, i.e., with an unprecedented precision of 1 part in 3 million\cite{novosibirsk}.  At CLEO, in a single measurement, $\mathcal{B}(J/\psi\to e^+e^-)/\mathcal{B}(J/\psi\to\mu^+\mu^-)=0.997\pm0.013$, lepton universiality is confirmed to $\sim1\%$\cite{cleodilep}.  In another measurment, isospin conservation (violation) is confirmed to $\sim2\%$ (0.5\%) by measuring $\psi'$ decay to $\pi^+\pi^-J/\psi$ versus $\pi^0\pi^0J/\psi$ (to $\pi^0J/\psi$ versus $\eta J/\psi$)\cite{cleopsitrans}.  In the bottomonium sector new measurements at CLEO\cite{cleodielec} of the leptonic branching ratios of $\Upsilon(1S,2S,3S)$ states with a precision of $1-2\%$ have resulted into up to 55\% changes in long--accepted values.  The precision result, $R\equiv \Gamma_{ee}(1S)/\Gamma_{ee}(2S)=2.19\pm0.03$ can be now used to validate the latest result of unquenched lattice calculation, $R=1.91\pm0.05$.  It appears that lattice is getting there.  Today $|b\bar{b}>$, tomorrow $|c\bar{c}>$?

Let me move on from improved precision to real discoveries.  In the charmonium sector, these have consisted of the discovery of the spin--singlet states $\eta_c'(2^1S_0,0^{-+})$ and $h_c(1^1P_1,1^{+-})$.  These bound states have long defied attempts to identify them, the main reson being that they are nearly impossible to populate via the vector $\psi'(2^3S_1,1^{--})$ formed in $e^+e^-$ annihilation.  Radiative decay to $\eta_c'$ is expected to be a very weak M1 transition, and that to $h_c$ is forbidden by charge conjugation.  Yet the importance of these spin--singlet states can not be exaggerated.  Together with the known spin--triplet states (or their centroid), these provide measures of the QCD hyperfine splitting, and therefore the $q\bar{q}$ \textbf{spin--spin interaction}.  So, our only hope of learning about the spin--spin interaction rests on the identification of these states of charmonium.

\subsection{The Discovery of the $\eta_c'(2^1S_0)$ State of Charmonium}

The only hyperfine splitting known in the heavy quarkonia is the mass difference of $1S$ charmonium states, $J/\psi$ and $\eta_c$:
$$\Delta M_{hf}(1S)=M(J/\psi)-M(\eta_c)=3096.9-(2980.4\pm1.2)=116.5\pm1.2~\mathrm{MeV}$$
A model--independent prediction for the hyperfine splitting of the $2S$ states is that
$$\Delta M_{hf}(2S)=M(\psi')-M(\eta_c')=\Delta M_{hf}(1S)\times\frac{M^2(\psi')}{M^2(J/\psi)}\times\frac{\Gamma(\psi'\to e^+e^-)}{\Gamma(J/\psi\to e^+e^-)}=68\pm2~\mathrm{MeV}$$

In 1982, the Crystal Ball Collaboration at SLAC claimed identification of $\eta_c'$ and reported $M(\eta_c')=3594\pm5~\mathrm{MeV}$, which would correspond to $\Delta M_{hf}(2S)=92\pm5$ MeV, about 35\% \textbf{larger} than expected.  Fortunately, or unfortunately, the Crystal Ball observation was never confirmed, and $\eta_c'$ remained unidentified for 30 years, until now.  

\begin{figure}[!tb]
\begin{center}
\includegraphics[width=2.5in]{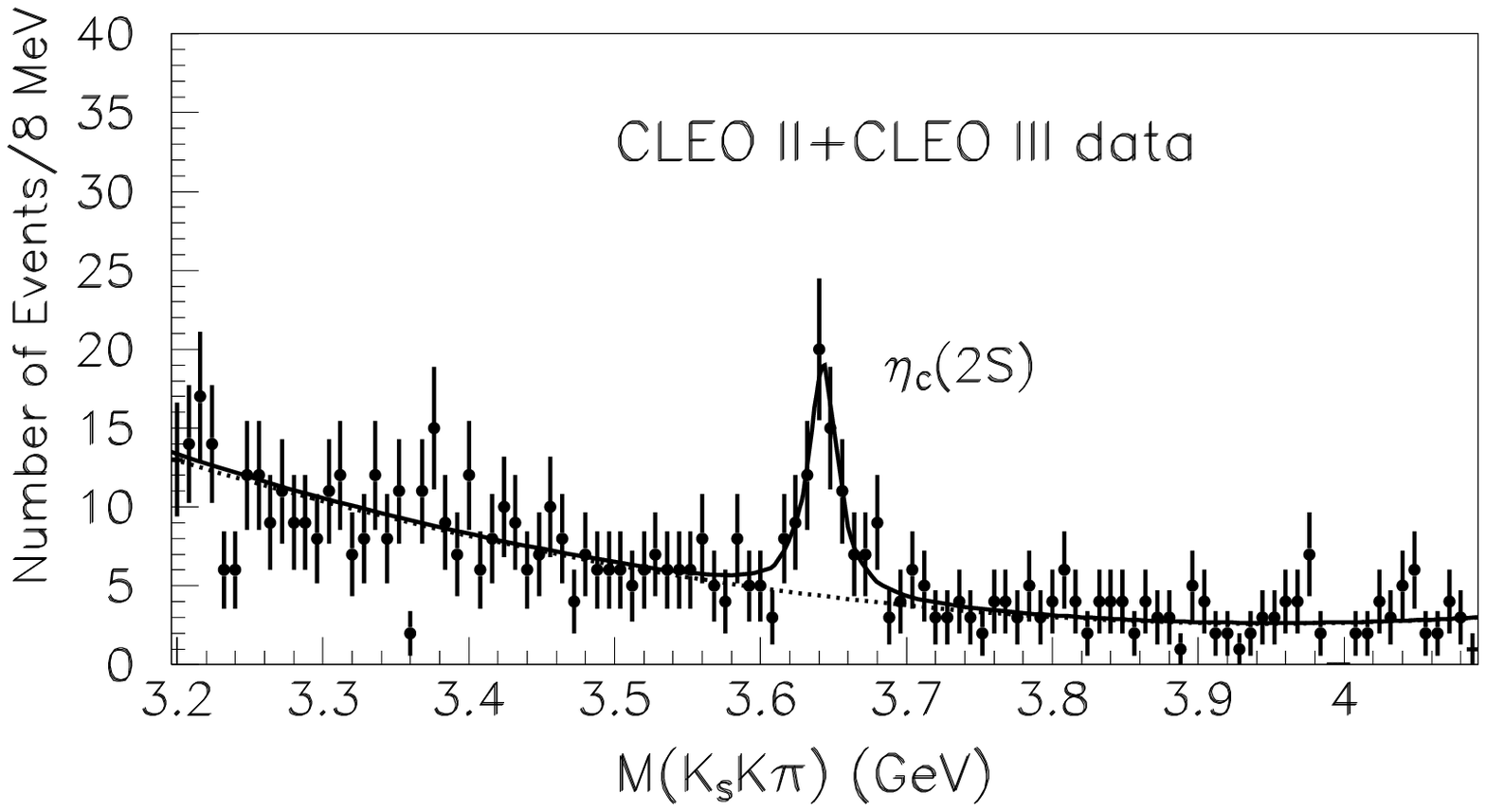}
\includegraphics[width=2.4in]{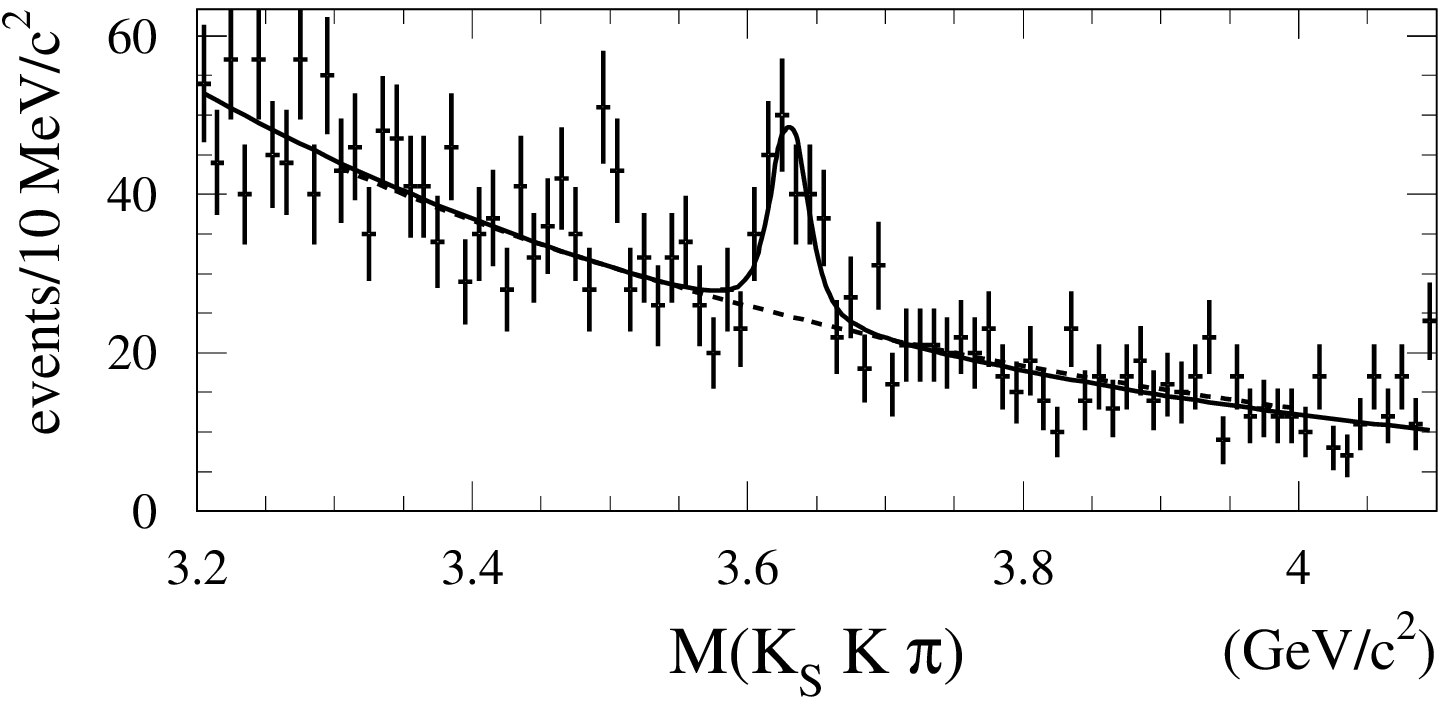}
\end{center}
\caption{The invariant mass $M(K_SK\pi)$ spectra from two--photon fusion measurements by CLEO (left) and BaBar (right).  The $\eta_c(2S)$ peak is prominent in both spectra.}
\end{figure}

In 2002, Belle claimed identification of $\eta_c'$ in the decay of 45 million $B$ mesons, $B\to K(K_SK\pi)$ and reported $M(\eta_c')=3654\pm10$ MeV, which would correspond to $\Delta M_{hf}(2S)=32\pm10$ MeV, a factor \textbf{two smaller} than expected and a factor \textbf{four} smaller than $\Delta M_{hf}(1S)$.  Could this be possible?  Well, perhaps!  There are two important ways $2S$ states differ from $1S$ states.  $1S$ states, with $r\approx0.4~\mathrm{f}$, lie in the Coulombic region (prop. to $1/r$) of the $q\bar{q}$ potential, $V=A/r+Br$, whereas the $2S$ states, with $r\approx0.8~\mathrm{f}$, lie in the confinement part (prop. to $r$) of the potential.  The spin--spin potential in the two regions could be different.  The second difference is that the $2S$ states, particularly $\psi'$, lie close to the $D\bar{D}$ breakup threshold at 3730 MeV, and can be expected to mix with the continuum as well as higher $1^{--}$ states.  All in all, it is important to nail down $\eta_c'$ experimentally, and measure its mass accurately.

This was successfully done by  CLEO\cite{cleoetacp} and BaBar\cite{babaretacp} in 2004 by observing $\eta_c'$ in two--photon fusion, $\gamma\gamma\to\eta_c'\to K_SK\pi$.  The two observations are shown in Fig.~4.  The average of all observations is $M(\eta_c')=3638.7\pm2.0$ MeV, which leads to $\Delta M_{hf}(2S)=47.4\pm2.0$ MeV, which is almost a factor \textbf{2.5 smaller} than $\Delta M_{hf}(1S)$.  Explaining this large difference is a challenge to the theorists.  The challenge for the experimentalists lies in completing the spectroscopy of $\eta_c'$, now that its mass is known.

\subsection{The Discovery of the $h_c(1^1P_1)$ State of Charmonium}

The interest in identifying $h_c$ is also related to the hyperfine or $\vec{s_1}\cdot\vec{s_2}$ interaction, but in a different way.  In the conventional $q\bar{q}$ potential, the confinement part is scalar, and therefore there is no long--range spin--spin interaction.  The $\vec{s_1}\cdot\vec{s_2}$ interaction is only due to the Coulombic part, and is a contact interaction proportional to the wave function at the origin.  In other words, it is finite only for $S$--states.  For $P$--wave, or higher $L$ states, the interaction is zero.  Thus, for $1P$ states, the $^1P_1$ state should be degenerate with the centroid of the $^3P_J$ states, or
$$\Delta M_{hf}(1P)\equiv \left<M(^3P_J)\right>-M(^1P_1)=0$$
There are many ways this prediction could be wrong, viz the simple way of calculating the centroid as $\left<M(^3P_J)\right>=[5M(^3P_2)+3M(^3P_1)+M(^3P_0)]/9$ could be (and is) wrong, or the $q\bar{q}$ potential could have a long--range $\vec{s_1}\cdot\vec{s_2}$ part in it.  It is therefore important to experimentally determine $\Delta M_{hf}(1P)$ by identifying the $h_c(1^1P_1)$ state of charmonium.  Numerous attempts to identify $h_c$ made during the last 35 years had failed.  Now, two attempts have succeeded.  The Fermilab E835 $\bar{p}p$ annihilation experiment\cite{e835hc} has claimed $h_c$ observation at $\sim3\sigma$ level\cite{cleohc} in the reaction $\bar{p}p\to h_c \to \gamma\eta_c$ and reported $\Delta M_{hf}(1P)=-0.4\pm0.2\pm0.2$ MeV.  CLEO has reported $h_c$ observation at $>6\sigma$ level in the reaction $\psi(2S)\to\pi^0h_c,~h_c\to\gamma\eta_c$.  Both inclusive ($\pi^0$ and $\gamma$ detected) and exclusive analyses were made, with the result $\Delta M_{hf}(1P)=+1.0\pm0.6\pm0.4$ MeV.  The exclusive spectrum is shown in Fig.~5. It appears that the $1P$ hyperfine splitting is very small, perhaps even zero.

\begin{figure}[!tb]
\begin{center}
\rotatebox{270}{\includegraphics[width=2.in]{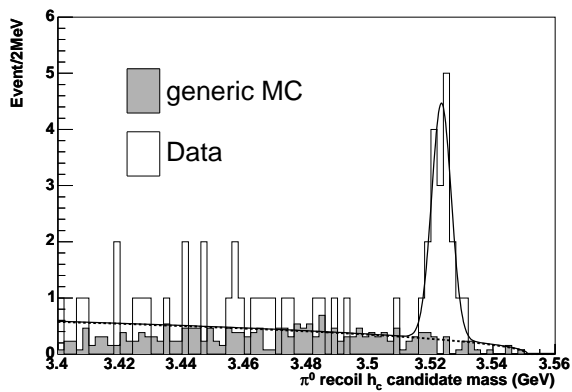}}
\end{center}
\caption{The spectra of $\pi^0$ recoil energies in the exclusive analysis of the reaction $\psi'\to\pi^0,~h_c\to\gamma\eta_c$, as measured by CLEO.  The peak corresponding to $M(h_c)$ is prominent.}
\end{figure}

With the discovery of $\eta_c'$ and $h_c$, a milestone has been reached in charmonium spectroscopy.  The bound state spectrum is now complete.

\section{New Results in Bottomonium Spectroscopy}

CLEO is now the only producer of $|b\bar{b}>$ bottomonium spectroscopy results.  Besides improving the precision of the branching ratio measurements for the radiative\cite{cleoupsrad} and leptonic \cite{cleodielec} decays of the Upsilon states, CLEO has made the first observation of a non--$\pi\pi$ transition between bottomonium states\cite{cleoupsomega}, and the first observation of a $D$--state in bottomonium\cite{cleoupsdstate}.  Since no bottomonium physics was presented at the Workshop, I will dwell no further on it, and return to meson spectroscopy below 5 GeV.

\section{The Unanticipated Discoveries}

The renaissance in hadron spectroscopy to which I referred earlier has been triggered by the discovery, or claimed discovery, of a number of unanticipated states.  This is very exciting, and also confusing, because the nature of these states is still not understood.  In chronological order, the series consists of the pentaquark (Spring8---Jan. 2003), $D_{sJ}$ (BaBar---March 2003), X(3872) (Belle---March 2003), X,Y,Z(3940) (Belle 2005), and Y(4260) (BaBar 2005).

\subsection{The Pentaquark}

There have been many reports of the sightings of the pentaquark $\Theta^+(1540)$ and its heavy quark siblings, and even a greater number of reported failures in finding them.  Google tells me that there were 123,000 entries for the pentaquark by March 28, 2006, 135,000 by April 10, and 151,000 by the time of this writing in July 25, 2006.  This tells you how exciting and intoxicating the idea is, or rather was.  
See, for example, Refs. 34 and 34.
The pentaquark has been on ``life--support'' for too long.  We should let it go---may it rest in peace!

\subsection{The Veteran X(3872)}

Discovered by Belle\cite{bellex}, and quickly confirmed by CDF\cite{cdfx}, D\O\cite{d0x}, and BaBar\cite{babarx}, the X(3872) state definitely exists.  Its average mass and width are $M(\mathrm{X})=3871.5\pm0.4~\mathrm{MeV}$, and $\Gamma(\mathrm{X})\le2.3~\mathrm{MeV}$, and it decays prominently in $\pi^+\pi^-J/\psi$.  What is debatable, however, is ``What is it?''  Numerous speculations have been made.  Is it a charmonium ($C=-$, $1^3D_{2,3}$ or $2^1P_1$) or ($C=+$, $1^1D_2$ or $2^3P_1$)\cite{xcharm}?  Is it a $C=+$ charmonium hybrid\cite{xhybrid}?  Is it a vector glueball mixed with vector charmonium\cite{xglueball}, or is it a $\overline{D^0}D^{0*}$ molecule\cite{xmolecule}?  A firm determination of the $J^{PC}$ of X(3872) would go a long way in sifting through  the above models.  Unfortunately, it is difficult to say that any of the claims made so far pin the $J^{PC}$ down unambiguously.  A number of reported decays, e.g., $\mathrm{X}\to\gamma J/\psi$, $\mathrm{X}\to \omega J/\psi$, $\mathrm{X}\to \overline{D^0}D^0\pi^0$, which would pin down some of the quantum numbers, suffer from limited statistics, and angular distribution studies and the dipion mass distribution in $\mathrm{X}\to J/\psi\pi^+\pi^-$ appear to be inconclusive.  The present toss--up appears to be between $J^{PC}=1^{++}$ and $2^{-+}$, with $1^{++}$ being a slight favorite.  With the goal of pinning down the binding energy of $\overline{D}D^*$ (if it is bound), CLEO has announced\cite{cleod0mass} a preliminary precision determination of the $D^0$ mass, $M(D^0)=1864.847\pm0.150\pm0.200~\mathrm{MeV}$, so that the best estimate of the binding energy of the $\overline{D}D^*$ molecule as $\mathrm{X}$ is now $E_b\equiv M(\overline{D^0}D^{0*})-M(\mathrm{X})=+0.31\pm0.64~\mathrm{MeV}$, which constrains the molecular model considerably more than before.  Further constraint would require a more precise measurement of the mass of X.


\begin{table}[!tb]
\tbl{Summary of Experimental data for the X, Y, Z states of Belle.}
{
\begin{tabular}{|c|c|c|c|c|c|c|}
\hline \hline
 & M (MeV) & N (evts) & $\Gamma$ (MeV) & Formed in/ & No decays to & Suggested? \\
 & & & & Decays to & & \\
\hline
X\cite{xxbelle} & 3936(14) & 266(63) & 39(26) & $e^+e^-\to J/\psi(\mathrm{X})$ & $\mathrm{X}\nrightarrow \overline{D}D$ & $\eta_c''(2^1S_0)$ \\
 & 3943(10) & 25(7) & 15(10) & $\mathrm{X}\to\overline{D}D^*$ & $\mathrm{X}\nrightarrow\omega J/\psi$  & \\
Y\cite{yybelle} & 3943(17) & 58(11) & 87(34) & $B\to K\mathrm{Y}$ & & $c\bar{c}$ hybrid \\
 & & &  & $\mathrm{Y}\to \omega J/\psi$ & $\mathrm{Y}\nrightarrow \overline{D}D^*$ &  \\
Z\cite{zzbelle} & 3929(5) & 64(18) & 29(10) & $\gamma\gamma\to\mathrm{Z}$ &  & $\chi_{c2}'(2^3P_2)$\\
 & & & & $\mathrm{Z}\to D\overline{D}$  & $\mathrm{Z}\nrightarrow \overline{D}D^*$ & \\
\hline \hline
\end{tabular}
}
\end{table}

\subsection{Ringing in Belle: The X, Y, Z States}

With huge luminosities ($>350$ fb$^{-1}$) available to it, Belle has been announcing the discovery of resonance after resonance.  By now we have X($3940\pm10$), Y($3943\pm17$), and Z($3929\pm5$)), with all masses being consistent with 3935 MeV.  The latest results for these states are summarixed in Table~I.  As shown in the table, the errors in the parameters are large enough, and the evidence at least for \textit{non-observations} is tentative enough so that not all three, X, Y, Z, need be distinct and separate states.  This is particularly true for X and Z whose widths are similar, and both of them could accomodate different levels of $D\overline{D}$ and $\overline{D}D^*$ decays.  It is obvious that lots more experimental data will be needed before the dust settles down.  In this respect, it is very disappointing that BaBar, with comparable data sets available to it, has maintained an ominous silence on X, Y, Z.  With the excitement these states have generated, one would think that BaBar would jump right in, and prove or disprove the Belle claims!

\begin{figure}[!tb]
\begin{center}
\includegraphics[width=2.3in]{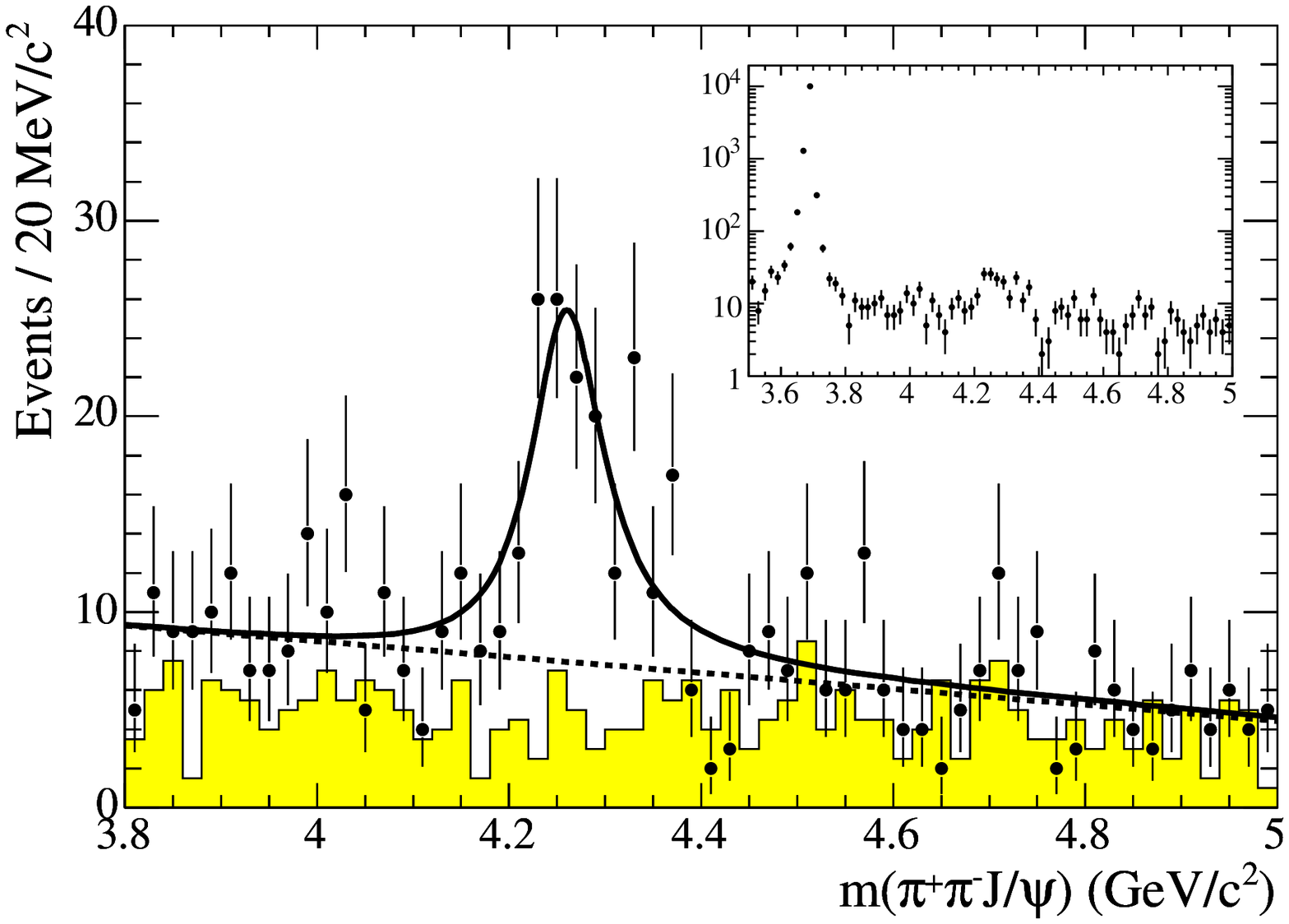}
\includegraphics[width=2.05in]{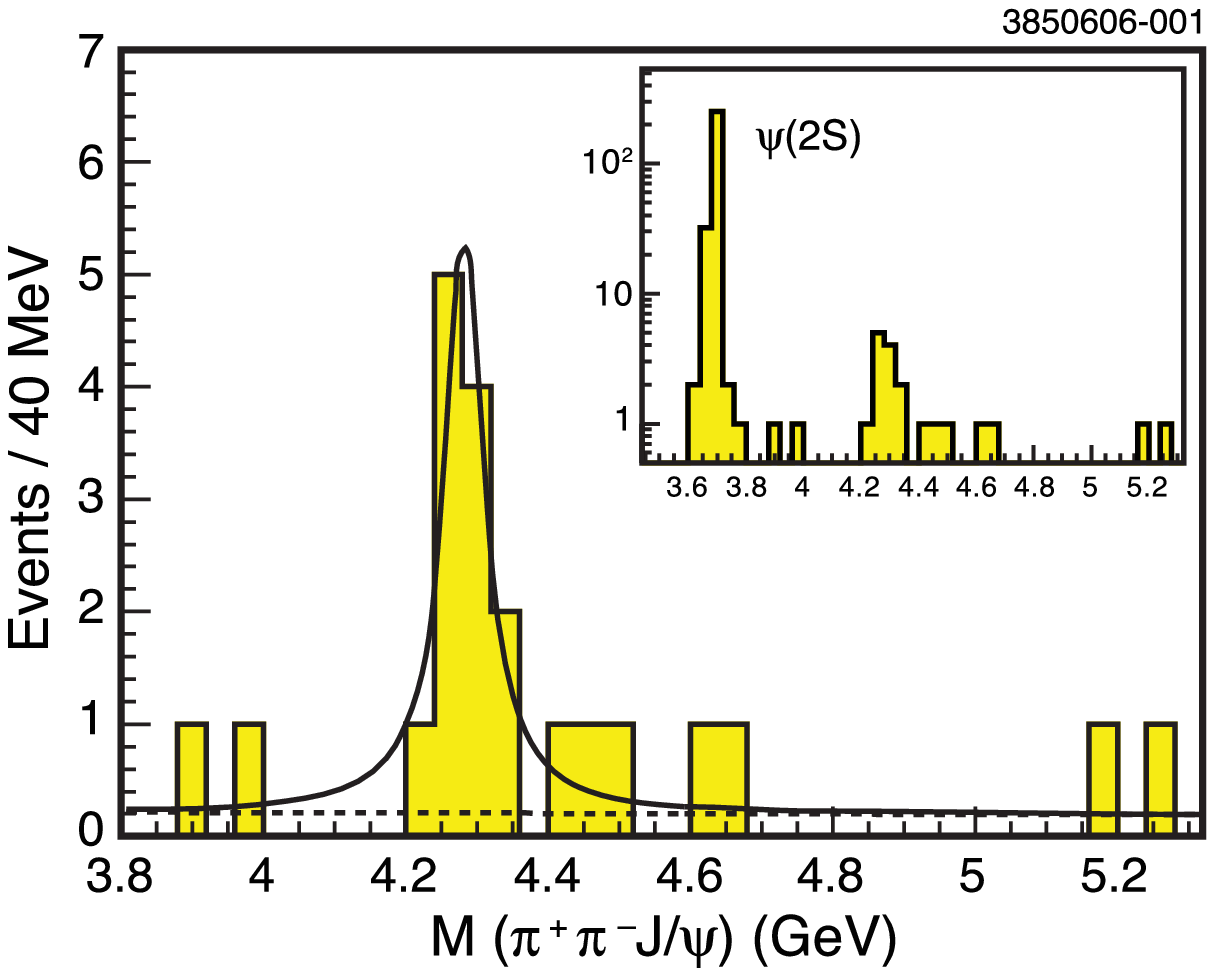}
\end{center}
\caption{The Y(4260) state observed in ISR production, and decay, $\mathrm{Y}\to\pi^+\pi^-J/\psi$; (left) by BaBar, (right) by CLEO.}
\end{figure}


\subsection{BaBar Chimes in: The Y(4260) State}

While BaBar has regretably stayed away from Belle's X, Y, Z states, it has exploited its expertise in ISR (initial state radiation), or RR (radiative return), studies to present another surprise\cite{babary}.  It has presented evidence for the observation of a broad resonance formed in $e^+e^-$ annihilation following ISR, and decaying into $\pi^+\pi^-J/\psi$ (see Fig.~6).  The parameters of the resonance are $M(\mathrm{Y})=4259\pm8^{+2}_{-6}~\mathrm{MeV}$, and $\Gamma(\mathrm{Y})=88\pm23^{+6}_{-4}~\mathrm{MeV}$.  The formation of $Y(4260)$ in ISR and the S--wave dipion mass distribution essentially ensure that $J^{PC}(\mathrm{Y}(4260))=1^{--}$.  This comes as a great surprise, because there is a deep minimum in the $R\equiv\sigma(e^+e^-\to\mathrm{hadrons})/\sigma(e^+e^-\to\mathrm{leptons})$ measurement at 4260 MeV, which would militate against the presence of a vector with this mass.  It is therefore extremely important to confirm BaBar's observation.  This has been done in two new measurements at CLEO.

In the first measurement, CLEO\cite{cleoy4260} has confirmed a strong (factor $\sim7$) enhancement of the $\pi\pi J/\psi$ yield at 4260 MeV compared to the yield at 4160 MeV and 4040 MeV.  Lest one argue that enhancement at a single energy point does not confirm a resonance, in a second measurement, CLEO has confirmed the Y(4260) in ISR, exactly the same measurement as BaBar (see Fig.~6).  They obtain $M(Y)=4283^{+17}_{-16}\pm4~\mathrm{MeV}$, and $\Gamma(Y)=70^{+40}_{-25}~\mathrm{MeV}$.  The measurement has very small background, and a very clean ($\sim5\sigma$) signal.  So, the Y(4260) $1^{--}$ state is very real.  But, what is it?  No charmonium vector is expected with this mass, which deepens the mystery.  The usual speculations, such as misplaced charmonium, a $c\bar{c}$ hybrid, a molecule, etc., abound.  Clearly, lots more measurements are needed to sift through these speculations. 

As a postscript to the discovery of all these new and unanticipated states, and the uncertainty about their nature, let me recall that the discovery of $J/\psi$\cite{jpsiexp} was followed by a spate of theoretical papers about its nature\cite{jpsitheory}, most of which, even by famous Nobelists, turned out to be wrong.  That is how physics develops. 

Let me now move to the non--quarkonium results presented at the workshop.

\section{Heavy--Light ($Q\bar{q},~\overline{Q}q$)}

Here the heavy quark can be either the charm or bottom quark, and the light quark can be $n(=u,d)$, or $s$.  The convention for labeling the quantum numbers of heavy--light mesons is to couple the spin $s_Q$ of the heavy quark with the $j_q=l+s_q$ of the light quark to give the total spin $J=j_q+s_Q$.

\subsection{Open--Charm, or $D(=c\bar{n},~c\bar{s})$ Mesons}

We have had considerable activity in this sector, with BaBar, Belle and CLEO (particularly in its reincarnation as CLEO-c) making important contributions.

The first exciting news was BaBar's discovery\cite{babarnewd} of a new narrow width $D^*_s(c\bar{s})$ meson with $M=2317~\mathrm{MeV}$, $J^P=0^+$, unexpectedly below the $DK$ threshold.  This was followed by the discovery of its $J^+=1^+$ partner with $M=2483~\mathrm{MeV}$, below the $D^*K$ threshold, by CLEO\cite{cleonewd}.  These mesons were expected to lie above their respective thresholds, and to be therefore quite wide.  Theoretical suggestions to explain the observed characteristics of these mesons again range from simple $|c\bar{s}>$ to $DK^*$ molecules, to $|c\bar{s}>$ mixed with $DK$ and $D^*K$ continua, and to tetraquark, with no resolution so far.  The corresponding non-strange $|c\bar{d}>$ states would be expected at 2217 MeV ($0^+$) and 2363 MeV ($1^+$).  These are expected to be wide, and have not been firmly identified so far.

Admittedly, the main interest in heavy--light mesons rests in the study of their weak interaction properties, and CLEO-c has made this their major goal.  Hopefully precision measurements in this sector will go a long way towards validating the lattice QCD calculations.  CLEO has reported preliminary form factor results, $f(D^+)=222.6\pm16.7\pm^{+2.8}_{-3.4}~\mathrm{MeV}$\cite{cleodff} and $f(D_s^+)=282\pm16\pm7~\mathrm{MeV}$\cite{cleodsff}.  The corresponding predicitions from unquenched lattice calculations are $f(D^+)=201\pm3\pm17~\mathrm{MeV}$ and $f(D^+_s)=249\pm3\pm16~\mathrm{MeV}$\cite{latticedff}.  CLEO has also measured a large number of hadronic and weak decays of $D$ and $D_s$ mesons.

\subsection{Open--Beauty, or $B(=b\bar{n},~b\bar{s},~b\bar{c}$ Mesons}

This is the domain of CDF and D\O~contributions.  The latest of their contributions relate to the identification and mass measurements of the excited $B$--mesons.  In the notation $J(j_q)$ the masses are:\\
\begin{tabular}{lll}
CDF: & $M(B_1^0)=5734\pm3\pm2$ MeV, & $M(B_2^{*0})=5738\pm5\pm1$ MeV, \\
D\O: & $M(B_1^0)=5721\pm3\pm5$ MeV, & $M(B_2^{*0})=5746\pm5\pm5$ MeV, \\
\end{tabular}
\\
and the tour-de-force measurement by CDF is $M(B_c)=6275.2\pm4.3\pm2.3~\mathrm{MeV}$,
which compares with the lattice prediction of $M(B_c)=6304\pm12^{+18}_{-0}~\mathrm{MeV}$.

\subsubsection{The \textbf{Triumph} of the $B$-Factories}

Since we had a presentation of the latest results on $CP$ violation, let me summarize what we know now.  The best values of the angles of the unitary triangle are ${\alpha=105^{+15}_{-9}}^\circ$, ${\beta=21.7^{+1.3}_{-1.2}}^\circ$, and $\gamma=65\pm20^\circ$.  One set of CKM matrix elements is: $|V_{ud}|=0.97363(26)$, $|V_{us}|=0.2252(18)$, $|V_{ub}|=0.004455(33)$, and $|V_{td}|/|V_{ts}|=0.205(15)$, and, of course, $\sin2\beta=0.685\pm0.032$.

\section{Epilogue}

As I mentioned in the beginning, hadron spectroscopy has had a renaissance during the last five years.  We have heard about exciting new discoveries at this Workshop, and the future promises many more to come.

On a personal note, I want to thank the organizers and the hosts of this Workshop for a very successful and productive meeting in this wonderful town.

This work was supported by the U.S. Department of Energy.

\end{document}